\documentclass[a4paper,fleqn,usenatbib]{mnras}

\usepackage{newtxtext,newtxmath}

\usepackage[T1]{fontenc}
\usepackage{ae,aecompl} 
\usepackage{xcolor}
 
\usepackage{graphicx}	
\usepackage{amsmath, bm}	
\usepackage{amssymb}	

\title[Electron firehose instabilities]{Firehose instabilities triggered by the solar wind 
suprathermal electrons}

\author[S.M.Shaaban et al.]{
S. M. Shaaban,$^{1,2,3}$\thanks{E-mail: shaaban.mohammed@kuleuven.be}
M. Lazar,$^{1,2}$
R. A. L{\'o}pez$^{1}$
H. Fichtner,$^{2}$
S. Poedts$^{1}$
\\
 $^{1}$Centre for Mathematical Plasma Astrophysics, KU Leuven, Celestijnenlaan 200B, B-3001 Leuven, Belgium.\\
$^{2}$Institut f\"ur Theoretische Physik, Lehrstuhl IV: Weltraum- und Astrophysik, Ruhr-Universit\"at Bochum, D-44780 Bochum, Germany.\\
$^{3}$Theoretical Physics Research Group, Physics Department, Faculty of Science, Mansoura University, 35516, Mansoura, Egypt.
}

\date{Accepted XXX. Received YYY; in original form ZZZ}

\pubyear{2018}

\begin{document}
\label{firstpage}
\pagerange{\pageref{firstpage}--\pageref{lastpage}}
\maketitle

\begin{abstract}
In collision-poor plasmas from space, e.g., solar wind, terrestrial magnetospheres, 
kinetic instabilities are expected to play a major role in constraining the temperature 
anisotropy of plasma particles, but a definitive answer can be given only after 
ascertaining their properties in these environments. Present study describes the 
full spectrum of electron firehose instabilities in the presence of suprathermal 
electron populations which are ubiquitous in space plasmas. Suprathermal electrons 
stimulate both the periodic and aperiodic branches, remarkable being the effects shown 
by the aperiodic mode propagating obliquely to the ambient magnetic field which markedly 
exceeds the growth rates of the parallel (periodic) branch reported recently in \citet[MNRAS 
464, 564]{Lazar2017a}. Derived exclusively in terms of the plasma parameters, the 
anisotropy thresholds of this instability are also lowered in the presence of suprathermal 
electrons, predicting an enhanced effectiveness in the solar wind conditions. 
These results may also be relevant in various other astrophysical contexts where the 
firehose instabilities involve, e.g., solar flares, sites of magnetic field reconnection, 
accretion flows or plasma jets leading to shocks and co-rotating interactions in 
heliosphere, interstellar medium and galaxy clusters.

\end{abstract}




\section{Introduction}

Preferential acceleration of charged particles along a guiding magnetic field is a 
common feature of any collision-poor plasmas expanding in our Universe, such as stellar winds, 
or plasma jets and accretion outflows \citep{McComas2007, Paesold2000, Drake2006, Guo2014}.
An important amount of free energy is therefore expected to accumulate in the magnetic field 
direction, leading to kinetic anisotropies of plasma particles, like temperature ($T$, or 
pressure $P=n k_B T$) anisotropies, e.g., $T_{\parallel}>T_{\perp}$, where $\parallel$ and $\perp$ 
denote directions with respect to the magnetic field.
In the heliosphere plasma is sufficiently dilute and kinetic anisotropies are easily triggered by the huge amount of energy released by the Sun via the more or less energetic outflows of charged particles. 
However, the observations unveil quasi-stable states with only small deviations from isotropy. 
Particle-particle collisions are inefficient at large heliocentric distances in the solar wind,
and any excess of temperature (or mean kinetic energy) in direction parallel to the magnetic field
($T_\parallel > T_\perp)$ is expected to be constrained by the selfgenerated instabilities. Of these,
firehose instabilities appear to be the most plausible candidates, and if driven by the 
anisotropic electrons with an idealized bi-Maxwellian distribution, the theory predicts
two highly contrasting branches of electron firehose instability (EFHI) \citep{Li2000, Gary2003,
Camporeale2008, Hellinger2014}: The periodic electron firehose (P-EFH) with a finite oscillation in time, 
i.e., $\Re(\omega) \ne 0$, also known as the nonresonant firehose branch \citep{Gary2003}, and 
the aperiodic electron firehose (A-EFH) with $\Re(\omega) = 0$, which propagates only obliquely 
to the ambient magnetic field (i.e., $k_\perp > k_\parallel$, in terms of wave-vector 
components). In this case the A-EFH develops faster, with maximum growth rates much higher than 
P-EFH \citep{Li2000, Gary2003, Camporeale2008, Hellinger2014}, and may play the main role in 
reducing, eventually, the free energy, and leading to enhanced fluctuations which may scatter 
the electrons and limit their anisotropy. It is also known that firehose instability 
may influence macroscopic plasma properties, like viscous heating and thermal conduction, with 
implications for plasma dynamics at the magnetic field reconnection sites in the heliosheath 
\citep{Schoeffler2011}, and at larger scales in intracluster medium and accretion disks plasmas 
\citep{Sharma2006}, and may cause disruptions in the large-scale plasma jets triggering 
radiative fields \citep{Subramanian2012}.

For conditions more typical to the solar wind, the observed distributions show deviations 
from a standard Maxwellian shape, especially due to suprathermal populations, which enhance 
the high-energy tails and are well described by the (bi-)Kappa distribution functions 
\citep{Pierrard2010}. One should thus expect that suprathermal electrons may 
contribute with an additional free energy, enhancing the unstable emissions.
Such an expectation has recently been confirmed by a preliminary study which shows that 
growth rates of electron firehose instability propagating parallel to the magnetic field 
($k = k_\parallel$) increase in the presence of suprathermal electrons \citep{Lazar2017a}. 
Here we characterize the full spectrum of firehose unstable modes under the influence of 
these suprathermal populaions, obviously, with a focus on the oblique propagation, where 
both branches of the P-EFH and A-EFH instabilities are present.

The dispersion formalism is briefly described in the next section, on the basis of the general 
dispersion tensor for a bi-Kappa distributed plasma, which is given explicitly in the Appendix.
The unstable firehose solutions are derived using an instability dispersion solver, named DSHARK 
and dedicated to plasmas with bi-Kappa components \citep{Astfalk2015, Astfalk2016}. The effects
of suprathermal electrons are outlined by contrasting with idealized solutions for bi-Maxwellian 
electrons. Choosing plasma parameters in the range of measurements in the solar wind enables us 
to compare with similar works \citep{Gary2003, Camporeale2008}, which do not take into account 
the effects of suprathermal electrons. Sections 2.1 and 2.2 describe in detail both branches 
of firehose instabilities, for a case study (parametrized in Table 1) which is representative 
for the effects of suprathermal electrons.  The fastest growing branch is also identified 
providing general instability conditions, exclusively, in terms of plasma parameters. Conclusions 
of the present study are summarized in section~3.
%
\section{Full spectrum of EFH instabilities} \label{sec:2}
The general linear dispersion relation for the electromagnetic modes propagating at an 
arbitrary angle $\theta$ with respect to the uniform magnetic field ($\bm{B}_0=B_0 \hat{e}_z$) in 
a bi-Kappa distributed plasma is given by \citep{Summers1994, Shaaban2018St}
\begin{align}
0 = \det D(\omega, k, \theta)\,,\label{eq:4}
\end{align}
where $D(\omega, k, \theta)$ is a $3\times3$ matrix, whose elements are functions 
of wave-frequency $\omega$, wave-number $k$, and angle $\theta$, thermal velocity 
components, and the power-index $\kappa$. Explicit definitions of these elements are 
given in Appendix~\ref{sec:appendix}. In a working frame co-moving with the solar wind the velocity distributions of the suprathermal electrons are described by the bi-Kappa distribution function 
%
\begin{equation}
  f_e= \frac{\pi ^{-3/2} }{\Theta_{e,\perp}^{2}\Theta_{e,\parallel}}
  \frac{\Gamma\left( \kappa +1\right) }{\kappa^{3/2}\Gamma \left( \kappa -1/2\right) }
  \left( 1+\frac{v_{\parallel }^{2}}{\kappa\; \Theta_{e, \parallel}^{2}}
  +\frac{v_{\perp}^{2}}{\kappa\; \Theta_{e, \perp}^{2}}\right)^{-\kappa -1}.\label{eq:1}
\end{equation}
where $\int d^{3}v f_e= 1$, $\kappa>3/2$ is the power-index, and $\Theta_{e,\parallel, 
\perp}$ are defined by the components of (kinetic) temperature \citep{Lazar2017a}
\begin{align}
T_{e,\parallel}^\kappa=\frac{2 \kappa}{2 \kappa-3}\frac{m_e}{2 k_B}\Theta_{e ,\parallel}^2  
\; \; \;\text{and}  \; \; \; T_{e,\perp}^\kappa=\frac{2 \kappa}{2 \kappa-3}\frac{m_e}{2k_B}
\Theta_{e ,\perp}^2, \label{eq:2}
\end{align}
assumed anisotropic, i.e., $T_{e,\parallel}^\kappa > T_{e,\perp}^\kappa$, with respect 
to the background magnetic field $\bm{B}_0$. The effects of protons are minimized by 
considering them isotropic and Maxwellian
\begin{equation}
f_{p}\left(v\right)=\frac{1}{\pi^{3/2}\; \Theta_{p}^{3}}
\exp \left(-\frac{v^2}{\Theta_p^2}\right)\,,  \label{eq:3}
\end{equation}
where $\Theta_p=\sqrt{2k_{B}T_p/m_p}$ is the proton thermal velocity.
\begin{figure}	\centering
\includegraphics[scale=0.52,trim={0 0.9cm 0 0},clip]{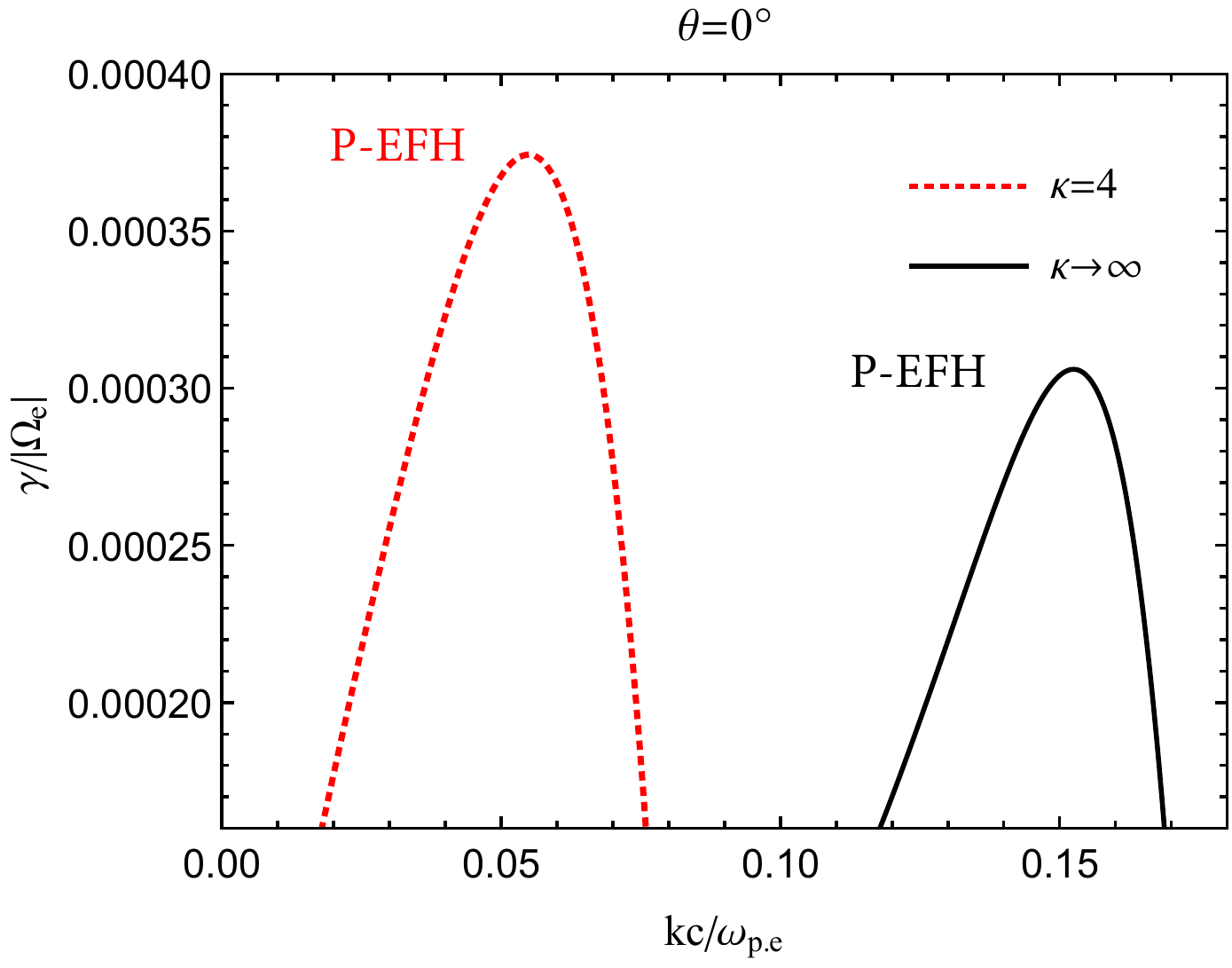}\\
\includegraphics[scale=0.5,trim={0 0 0 0},clip]{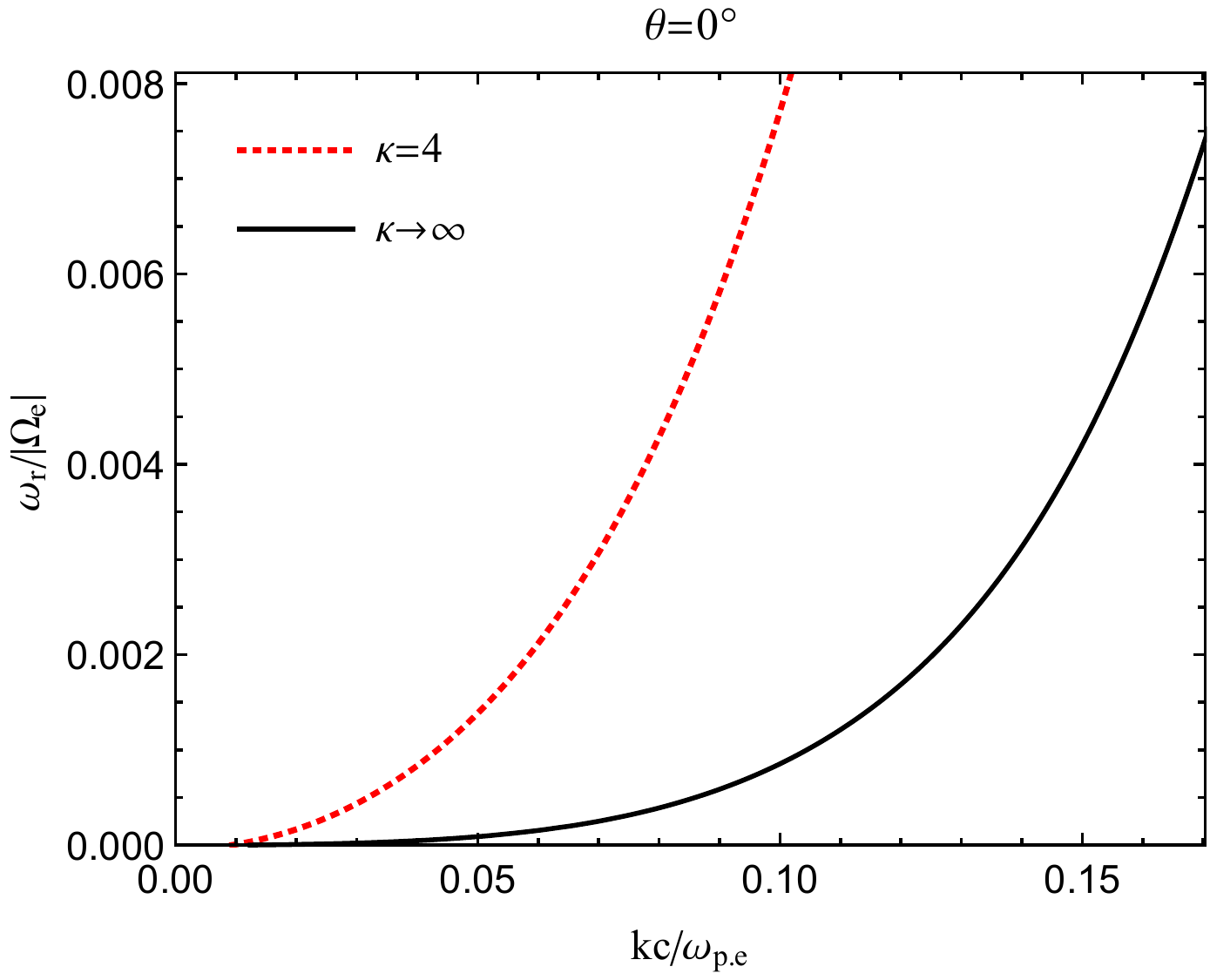}
\caption{Growth rates (top) and wave frequencies (bottom) of the P-EFH 
instability at $\theta=0^\circ$ and different electron power-index 
$\kappa=4$ (red dotted lines), $\kappa\rightarrow \infty$ (black solid lines). 
Growth rates and wave frequencies are normalized to the electron parameters.}\label{f1}
\end{figure}

Numerical analysis of the firehose unstable solutions is performed using DSHARK solver 
\citep{Astfalk2015}, recently developed to resolve the dispersion relation (\ref{eq:4}) for 
modes propagating at arbitrary angles with respect to the background magnetic field. 
Plasma parameters are tabulated in Table \ref{t1}, unless elsewhere provided. The effects 
of suprathermal electrons are outlined by contrasting the results obtained for a 
power-index $\kappa=4$ with those for a bi-Maxwellian limit ($\kappa\rightarrow \infty$) 
approaching the thermal core population of lower temperatures  \citep{Lazar2017a}
\begin{align}
T_{\parallel,\perp} < T_{\parallel,\perp}^\kappa=\dfrac{2\kappa}{2\kappa-3} T_{\parallel,\perp}, \,\,\,
{\rm or} \,\,\, \beta_{\parallel,\perp} < \beta_{\parallel,\perp}^\kappa= \dfrac{2\kappa}{2\kappa-3}
\beta_{\parallel,\perp},\label{eq:5}
\end{align}
where the plasma beta parameter $\beta \equiv 8\pi n_e k_B T/B_0^2$.

\subsection{Numerical solutions}

%
\begin{figure} 	\centering
\includegraphics[scale=0.53,trim={0 0.9cm 0 0},clip]{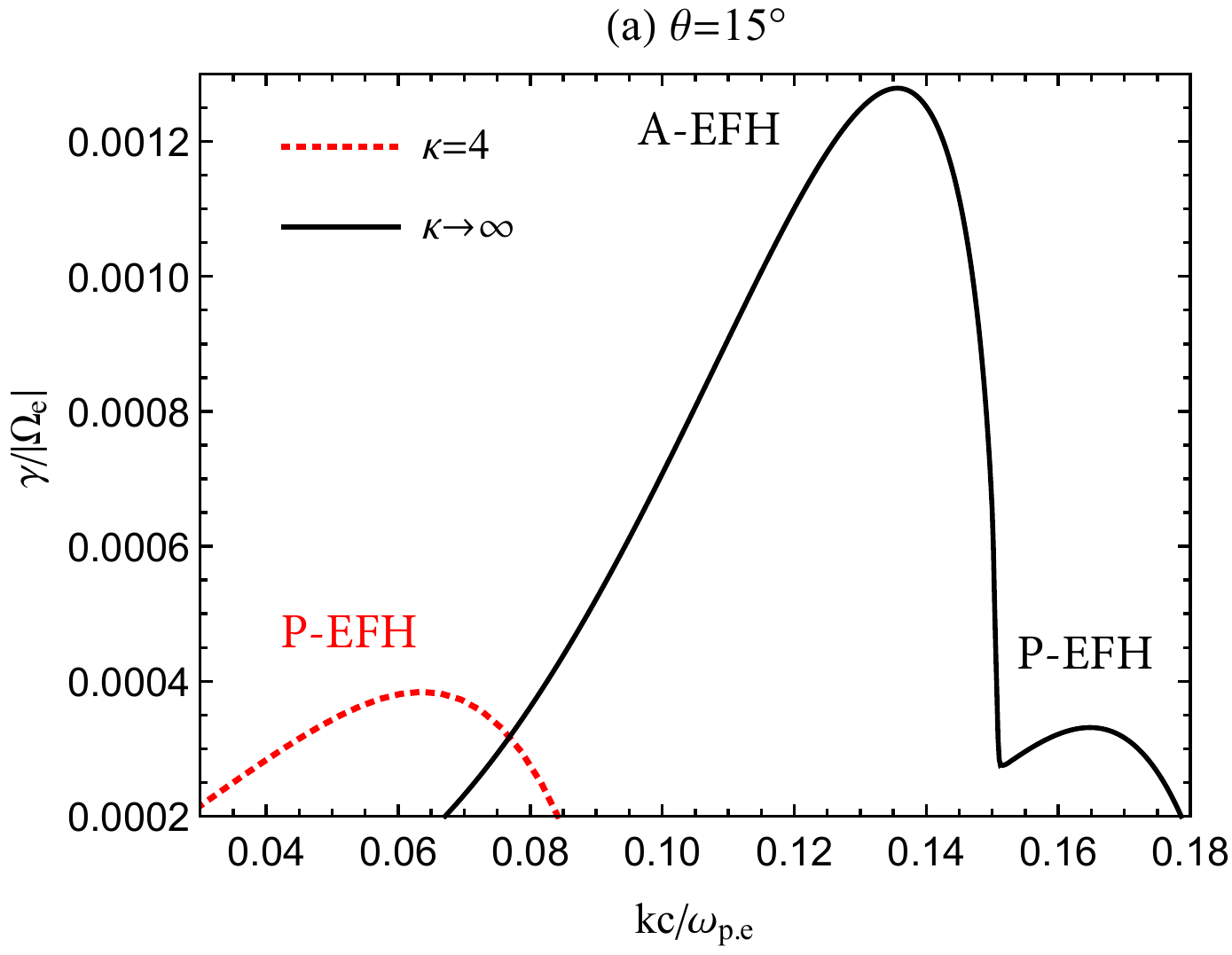} \\ \vspace{0.2cm}
\includegraphics[scale=0.53,trim={0 0.9cm 0 0},clip]{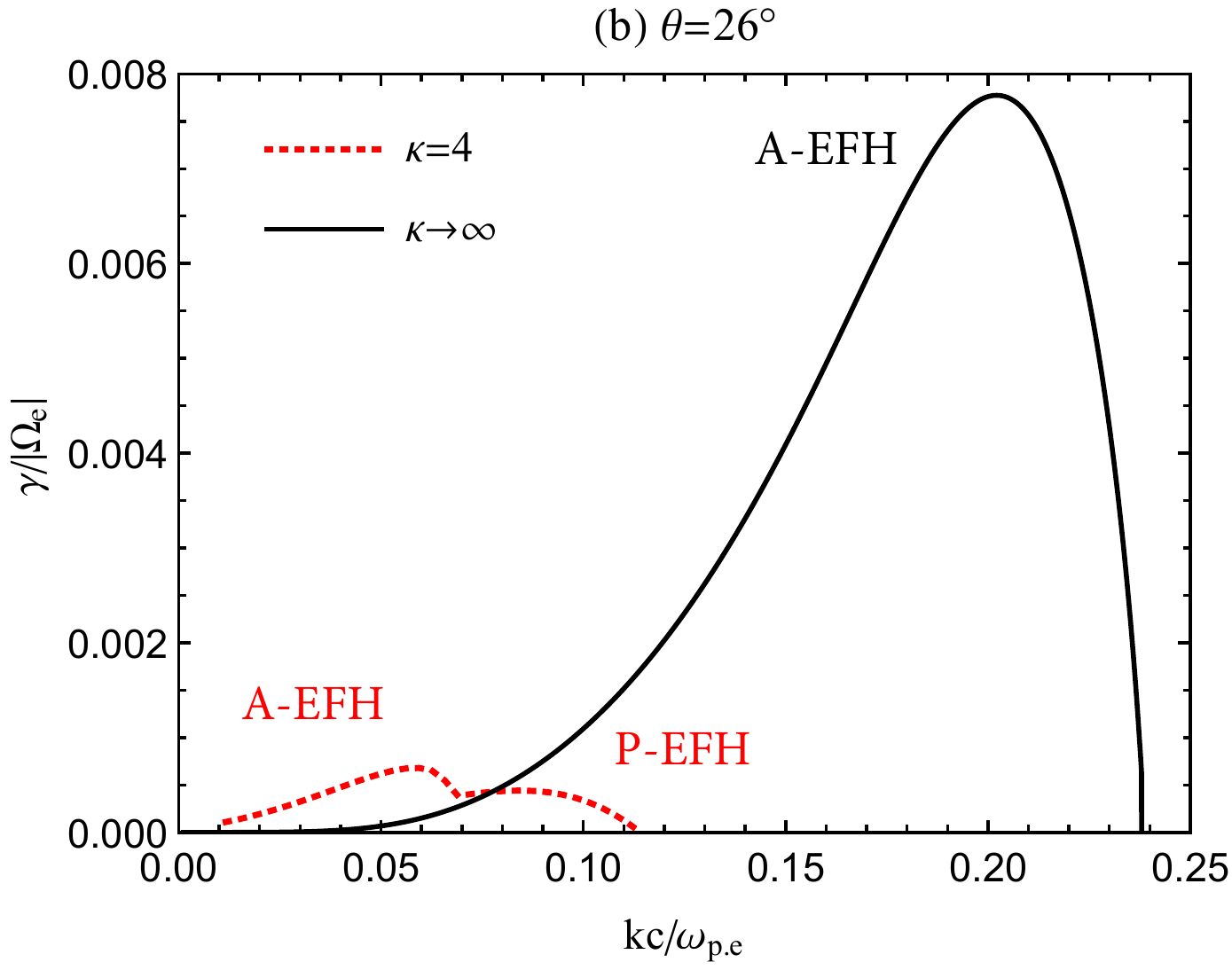}\\ \vspace{0.2cm}
\includegraphics[scale=0.5,trim={0 0.9cm 0 0},clip]{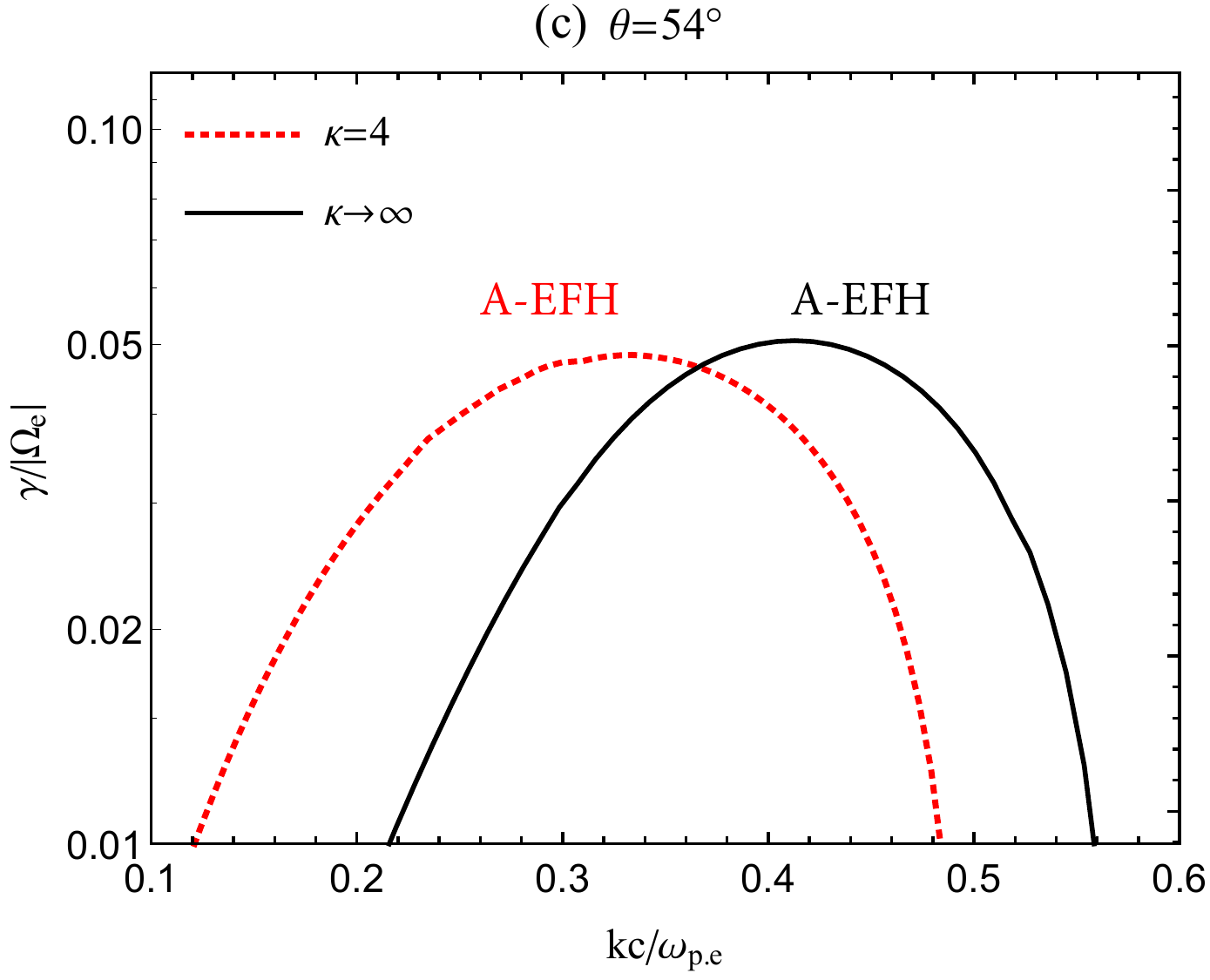}\\ \vspace{0.2cm}
\includegraphics[scale=0.5,trim={0 0 0 0},clip]{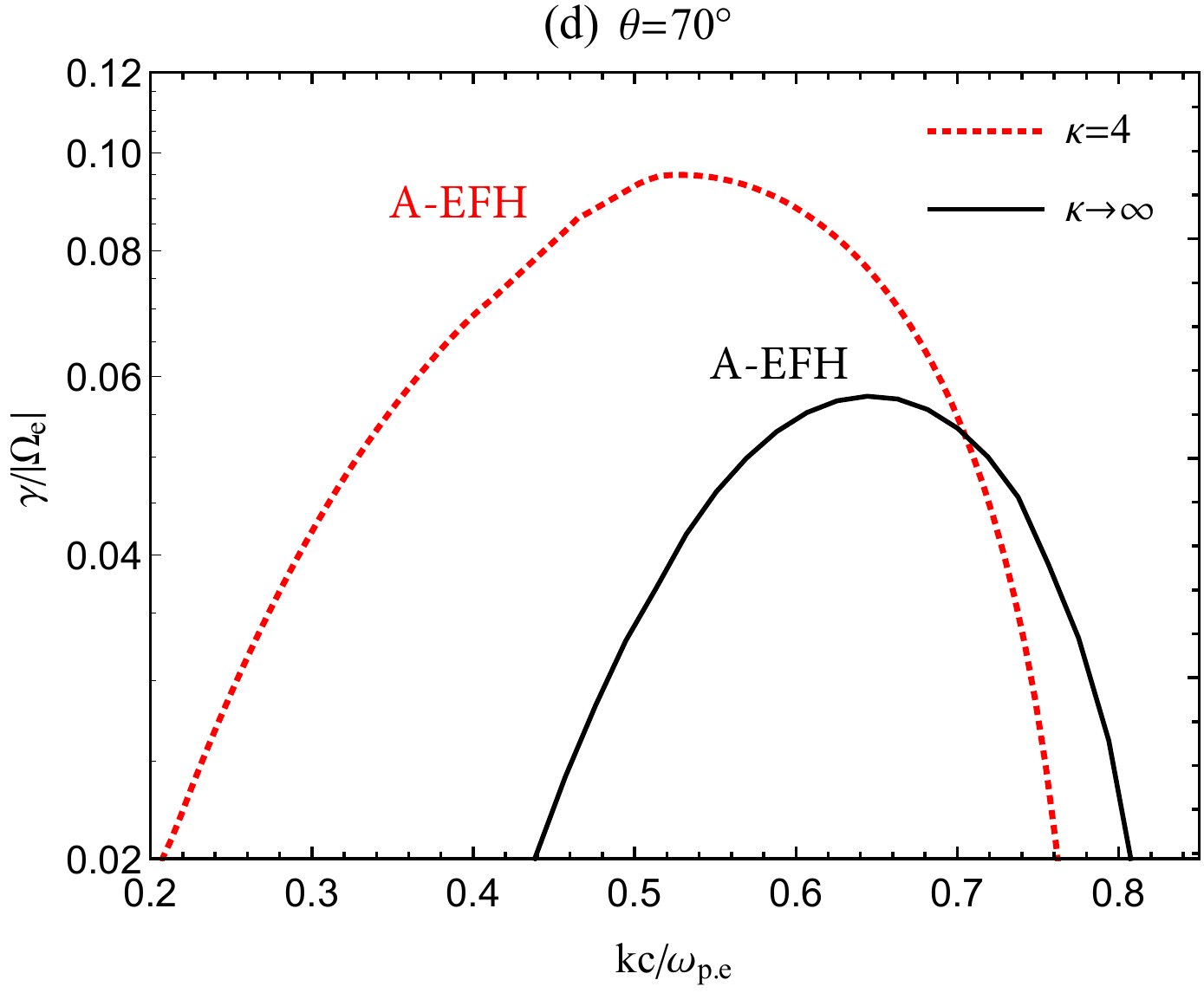}
\caption{Growth rates of the EFH instabilities for various angles of 
propagation $\theta$ and different electron power-index $\kappa=4$ 
(red dotted lines), $\kappa\rightarrow \infty$ (black solid lines).}\label{f2}
\end{figure}
\begin{figure} 	\centering
\includegraphics[scale=0.51,trim={0 0.9cm 0 0},clip]{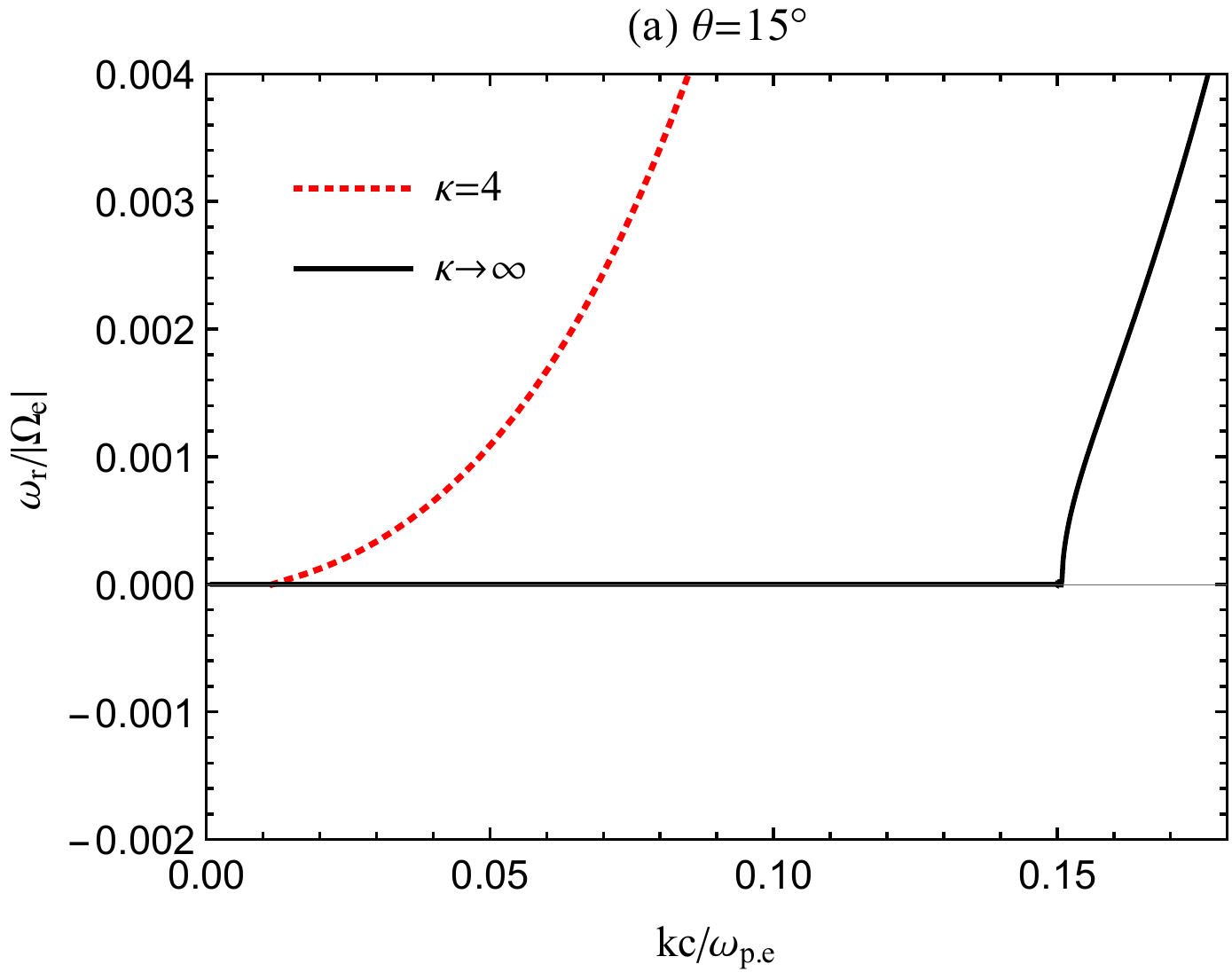}	\\ \vspace{0.2cm}
\includegraphics[scale=0.51,trim={0 0.9cm 0 0},clip]{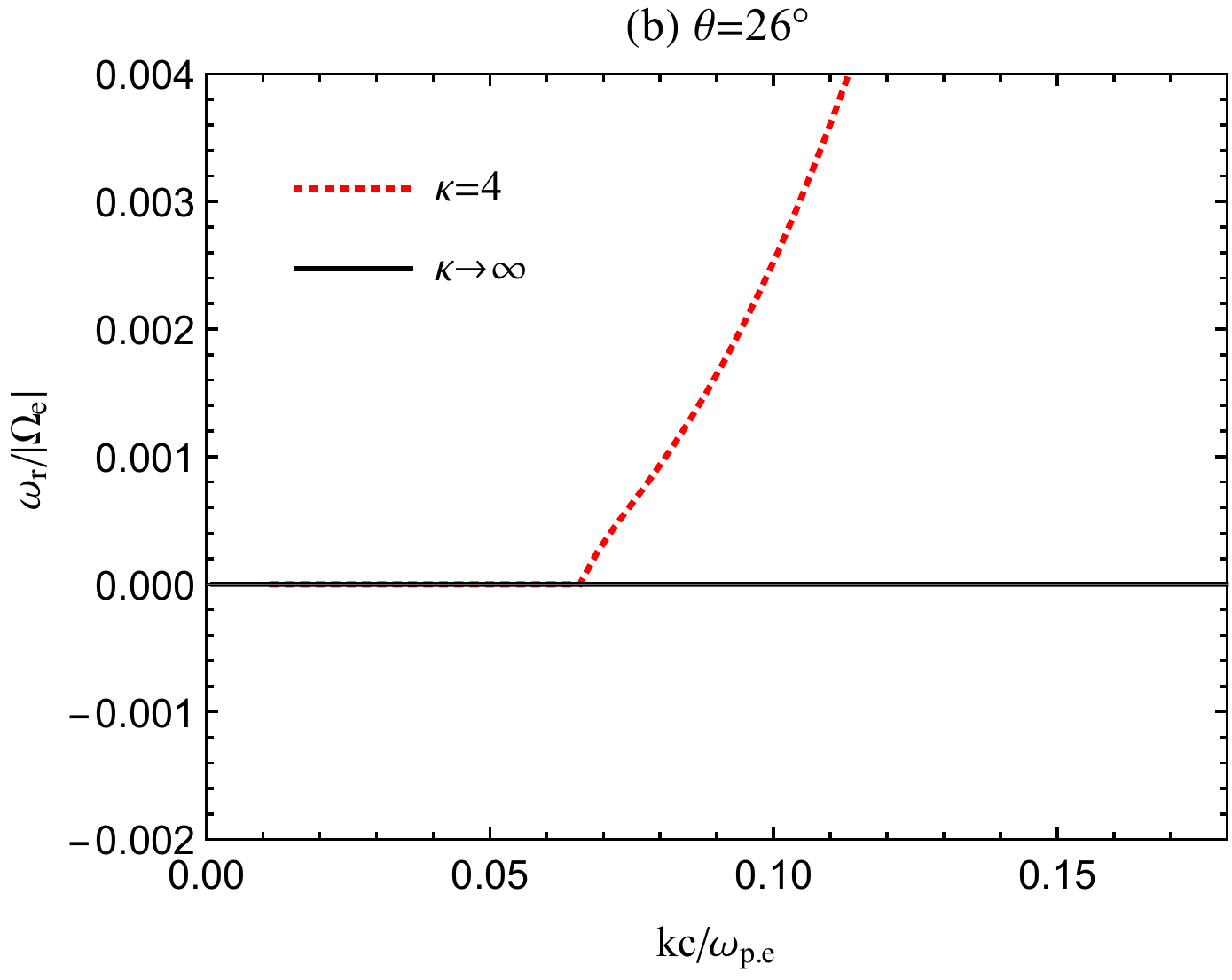}\\ \vspace{0.2cm}
\includegraphics[scale=0.53,trim={0 0.9cm 0 0},clip]{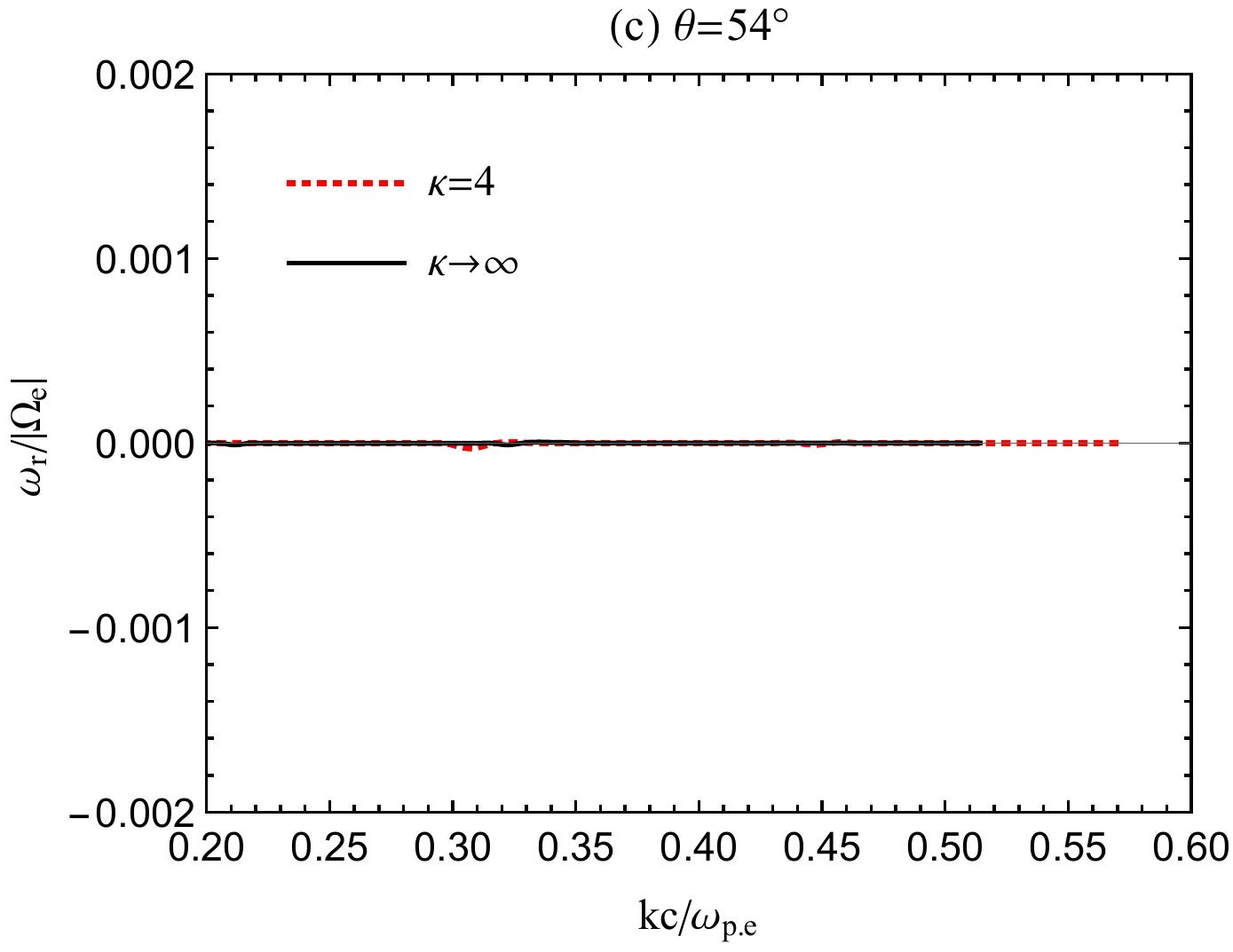}\\ \vspace{0.2cm}
\includegraphics[scale=0.52]{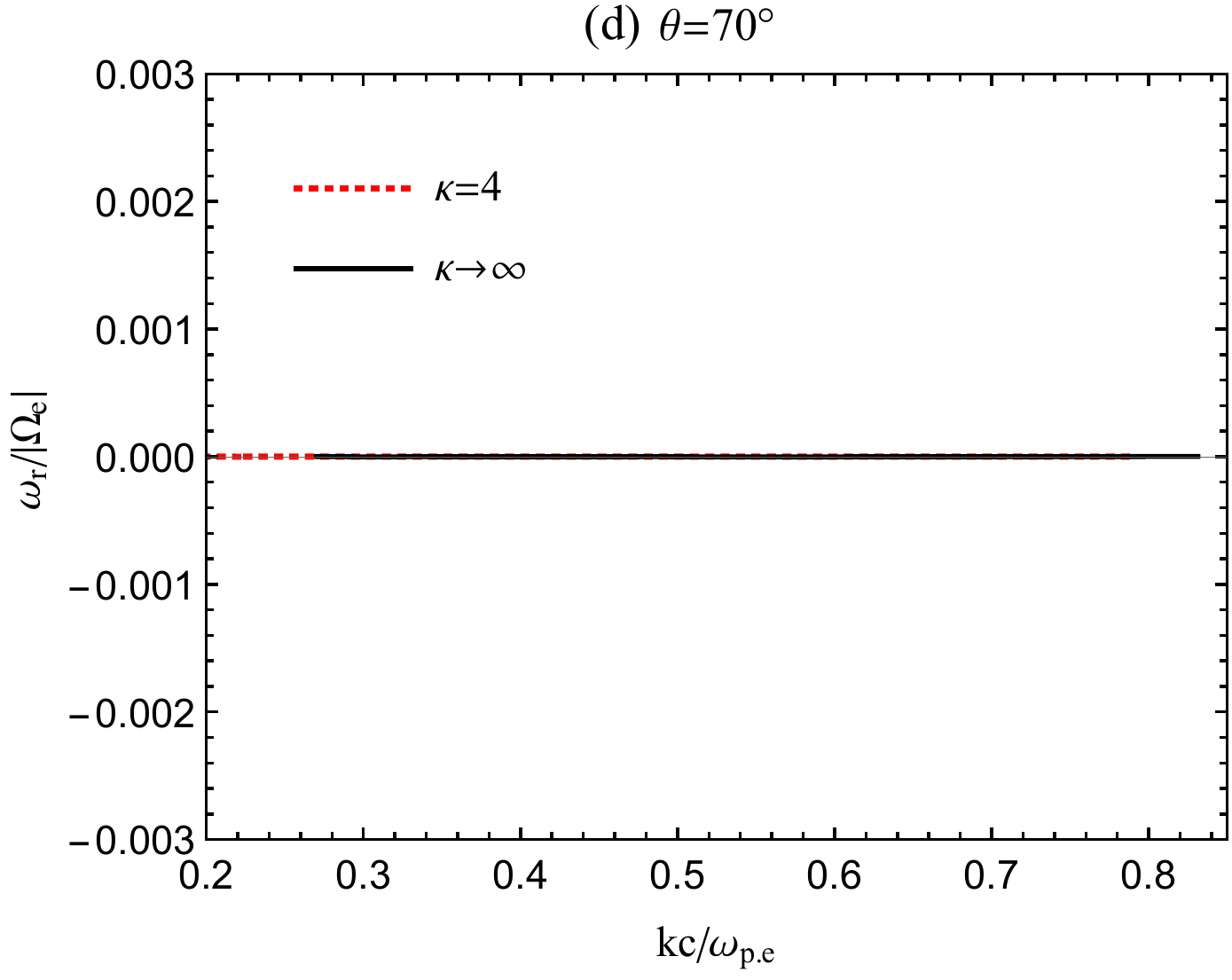}
\caption{The corresponding wave frequencies of the EFH instabilities 
growth rates in Figure \ref{f2}.}\label{f3}
\end{figure}

In order to build a comprehensive picture for the effects of suprathermal electrons
we start with a comparative analysis of P-EFH and A-EFH branches, e.g., in Figures~\ref{f1}-\ref{f3}, 
for the plasma conditions in Table \ref{t1} and various angles of propagation. 
The test case in Figure \ref{f1} ($A_e=0.5$ and $\beta_{e,\parallel}=4$) presents the 
dispersive and stability properties of the P-EFH instability in direction parallel to the 
magnetic field, i.e., $\theta=0^\circ$, and confirms the validity of a new normalization 
introduced in DSHARK with respect to the electron (scaling) parameters (replacing the 
original normalization to ion parameters). Both the maximum growth-rate and 
wave-frequency of the P-EFH instability increase in the presence of suprathermal electrons 
(red-dotted lines), i.e. for finite (low) values of $\kappa$, and the instability becomes 
more operative at lower wave-numbers. These results obtained with DSHARK are in perfect 
agreement with the P-EFH solutions provided by \citet{Lazar2017a}. 

The unstable solutions in Figures \ref{f2} and \ref{f3}, displaying, respectively, the 
growth rates and the corresponding wave frequencies, enable us to examine the effect of 
suprathermal electrons on the oblique modes, i.e., for $\theta=15^\circ$, $26^\circ$, 
$54^\circ$, and $70^\circ$. These angles of propagation are carefully chosen 
to highlight the most important effects of suprathermals on the EFH instabilities.
At small angles, i.e., $\theta=15^\circ, 26^\circ$, both P-EFH and A-EFH branches are 
present with modest growth-rates (Figure \ref{f2}), but restrain to lower wave numbers 
in the presence of suprathermals (red-dotted lines). Stimulation of P-EFH instability by 
the suprathermals, reported by \citet{Lazar2017a} for parallel propagation ($\theta= 0^\circ$),
is also observed for the oblique angles, e.g., in Figure \ref{f2}, panel (a), where both 
P-EFH peaks are present. For $\theta=26^\circ$ P-EFH unstable modes are still present 
only in the presence of suprathermals (black-solid line), when growth rates in panel (b) 
display two distinct peaks of A-EFH and P-EFH at low and large wave-numbers, respectively. 
This branch disappears with increasing $\theta$, and the peak of A-EFH instability significantly 
increases for a highly oblique propagation, e.g., for $\theta > 50^\circ$ in panels (c) and 
(d). The fastest growing modes, corresponding to the peaking growth rates at these angles, 
are markedly stimulated by the suprathermal electrons.
The corresponding wave frequencies are displayed in Figure \ref{f3}, enabling to identify 
and distinguish between different branches of EFH solutions. P-EFH unstable modes are 
LH polarized with $\Re(\omega) > 0$, while A-EFH modes have zero real frequency $\Re(\omega) 
= 0$. The wave frequency confirms the effect of suprathermal (energetic) electrons, in the 
presence of which the range of unstable wave-numbers translates to markedly lower values, 
for both instabilities, P-EFH at small angles, and A-EFH at highly oblique angles. 

\begin{figure}
	\centering
\includegraphics[scale=0.585]{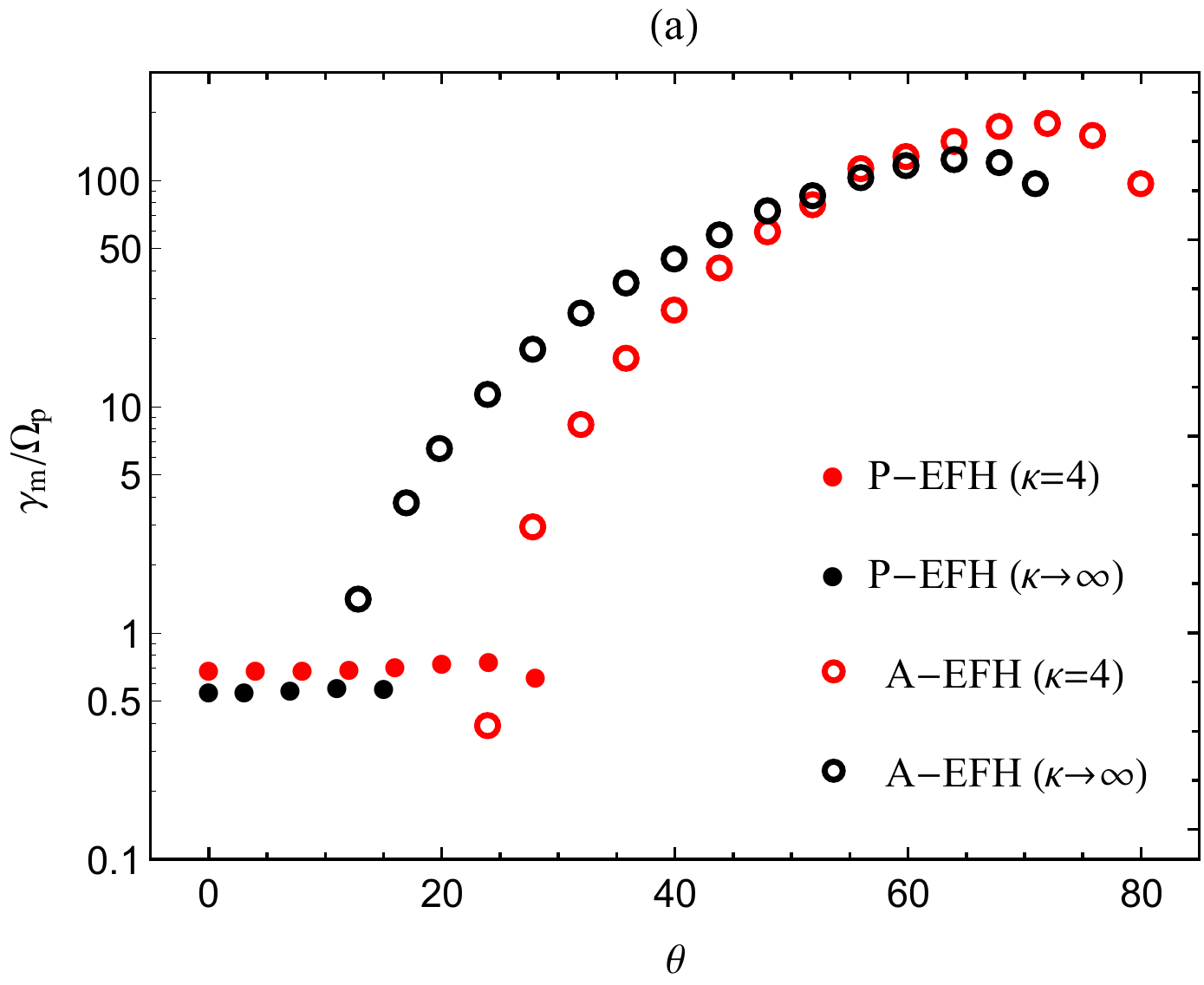}
\includegraphics[scale=0.585]{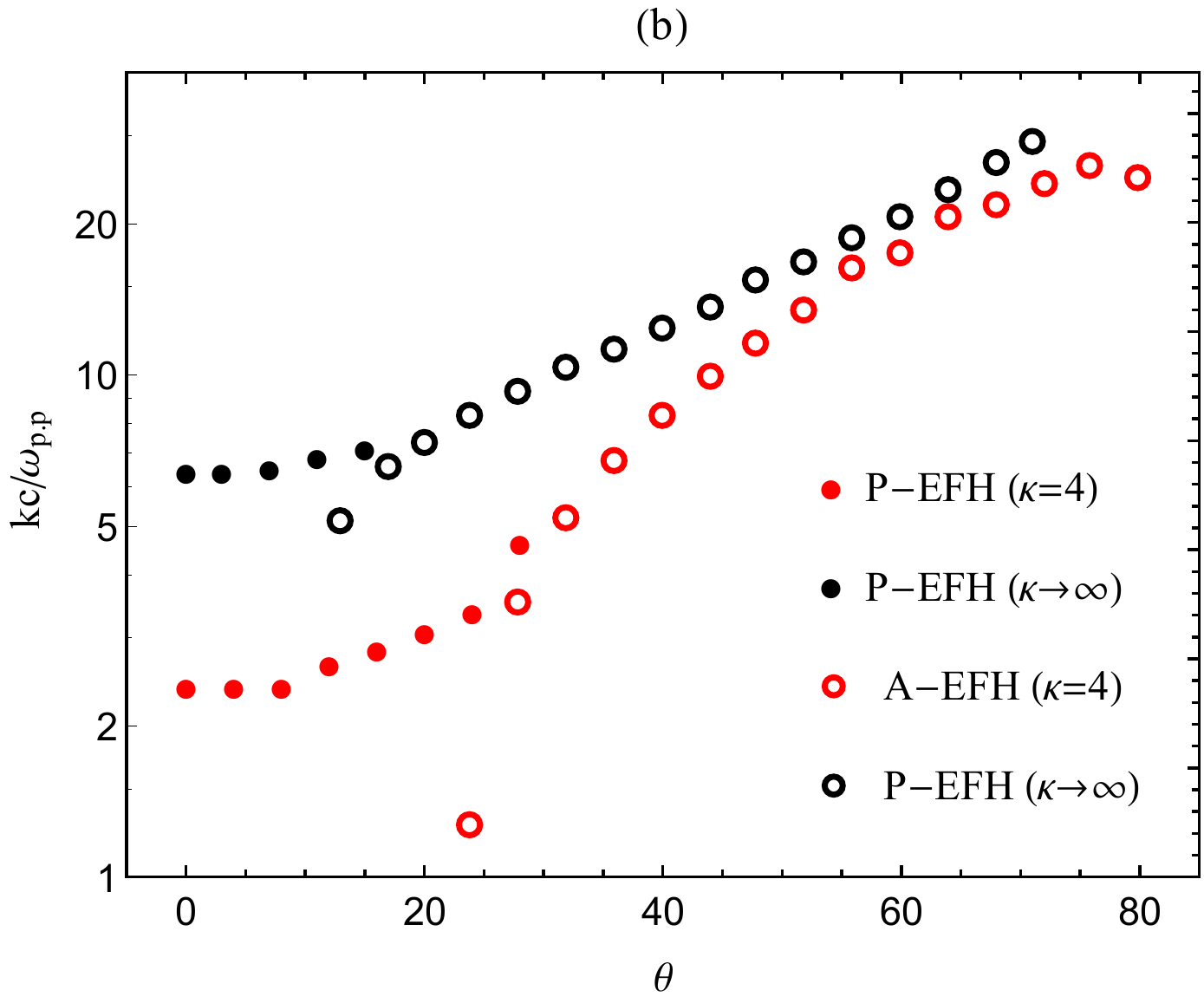}
\caption{Maximum growth rates, panel (a), and the corresponding maximum wave-number, panel (b), 
of P-EFH (solid circles) and A-EFH (open circles), for bi-Kappa ($\kappa=4$, red) and bi-Maxwellian distributed
electrons ($\kappa\rightarrow \infty$, blue). We normalized to the proton parameters for a better 
visualization.}\label{f4}
\end{figure}
\begin{figure}
\centering
\includegraphics[scale=0.5]{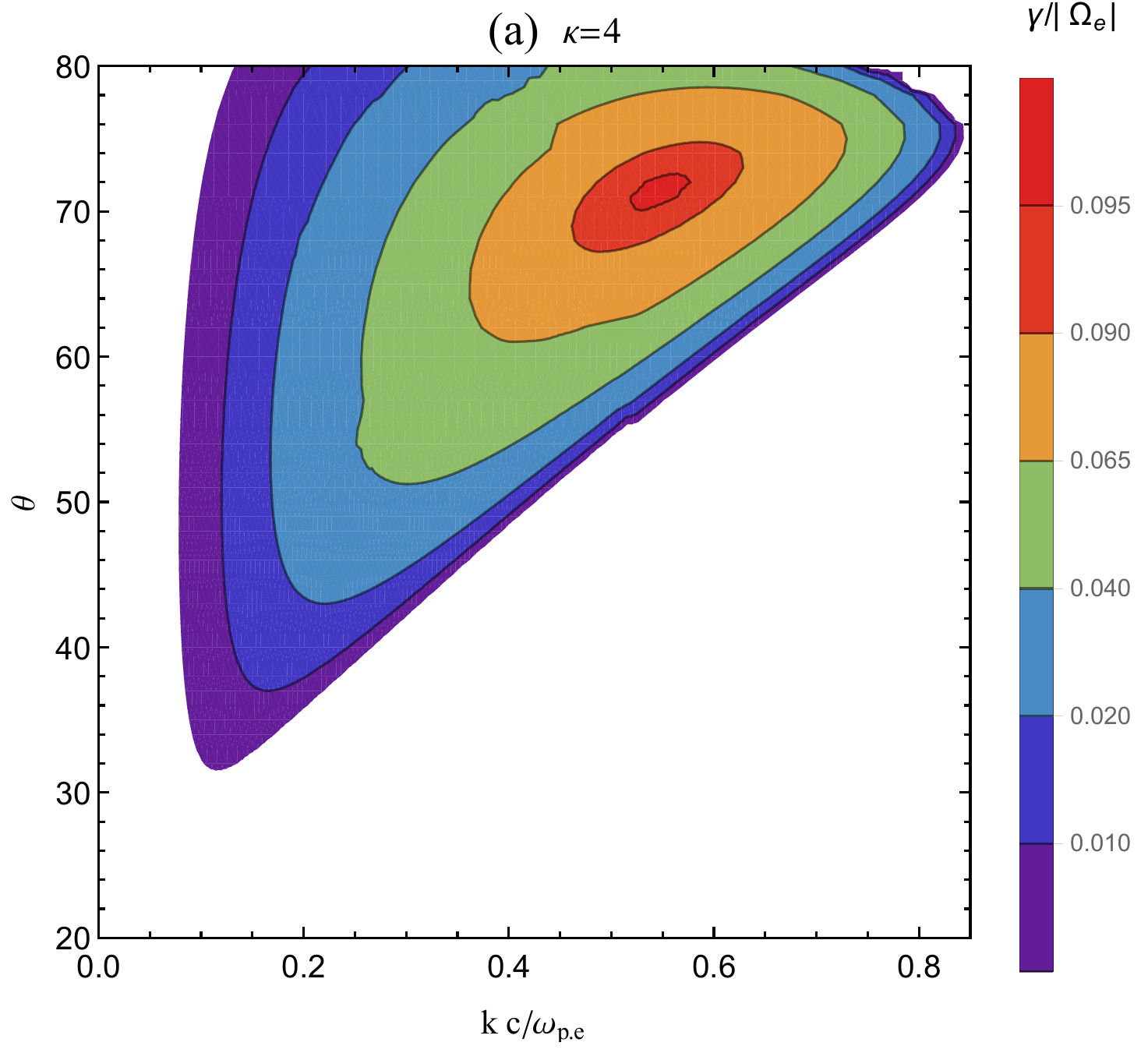}
\includegraphics[scale=0.44]{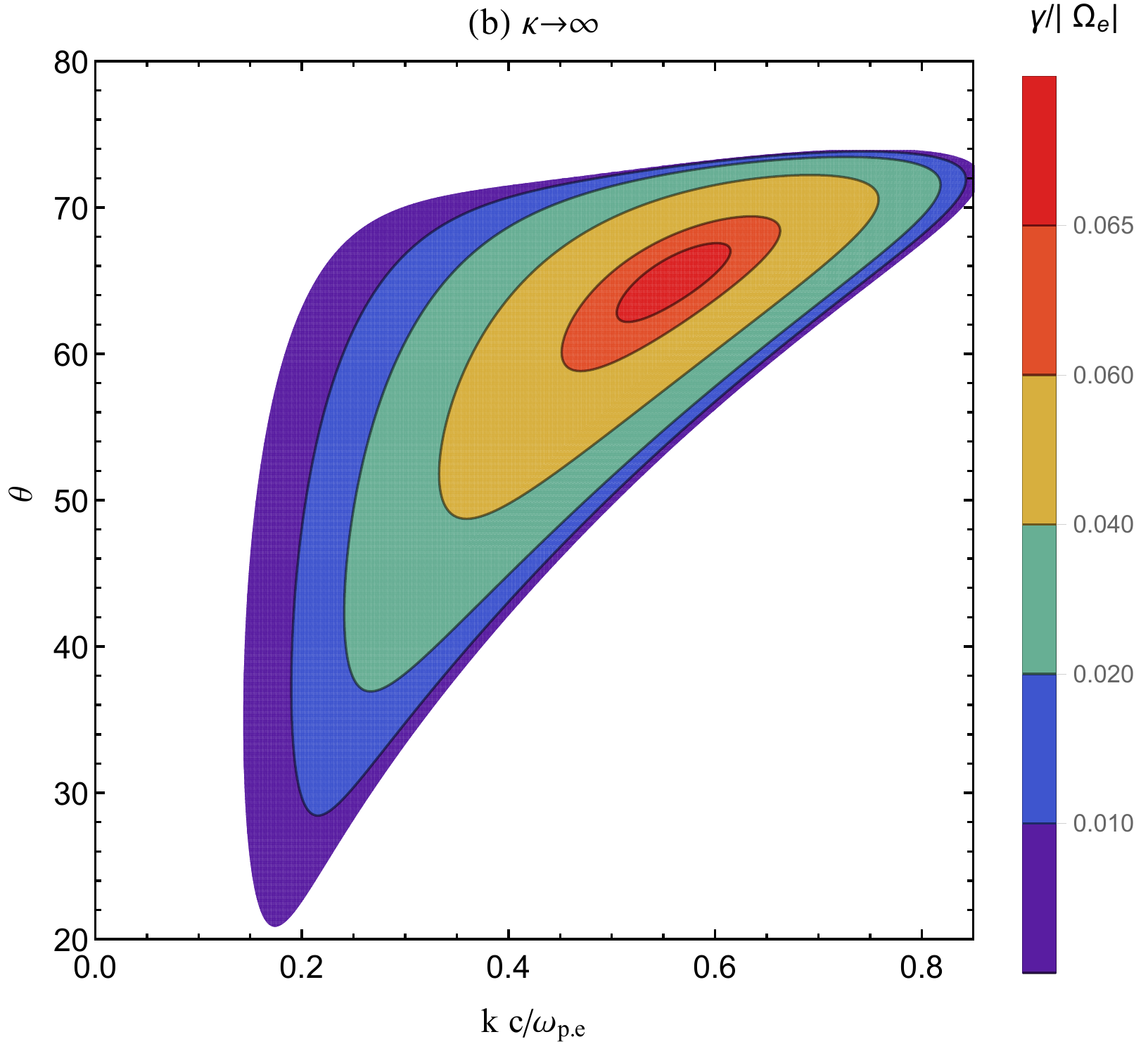}
\caption{3D representation of the A-EFH maximum growth rates as a function of $\theta$ and 
$kc/\omega_{p,e}$ for $\kappa=4$, panel (a) and $\kappa\rightarrow \infty$, panel (b). }\label{f5}
\end{figure}

Maximum growth rates $\gamma_m/\Omega_p$ and the corresponding wave-numbers $k_mc/\omega_{p,p}$ 
are plotted in Figure \ref{f4}, panels (a) and (b), respectively, as a function of $\theta$. 
P-EFH instability is indicated with solid circles, while open circles represent A-EFH instability. 
Standard Maxwellian ($\kappa\rightarrow\infty$) solutions (black) show profiles similar to 
those obtained in \citet{Paesold2000}, e.g., growth rates of A-EFH mode exceed the P-EFH 
mode at about $\theta \simeq 10^\circ$; a non-uniform variation of maximum growth rates which 
increase and then decrease with increasing angle of propagation, reaching the peak at about 
$65^\circ$ (also in agreement with the results in \citet{Maneva2016}). Solutions obtained under 
the influence of suprathermal electrons ($\kappa=4$, red) show that A-EFH branch 
becomes faster than the P-EFH at about $30^\circ$, and undergo a similar non-uniform variation 
with $\theta$, reaching its peak at about $72^\circ$. We can already point out two results: 
(i) for both P-EFH and A-EFH branches the peaks at oblique angles become more pronounced in 
the presence of suprathermals, and are obtained for slightly higher angles; (ii) A-EFH remains 
the fastest growing mode, and becomes even faster under the influence of suprathermals.
The wave-numbers $k_mc/\omega_{p,p}$ corresponding to maximum growth rates in panel (b)
increase with increasing $\theta$, and show that, whenever the growth rates display two peaks,
i.e., for P-EFH and A-EFH branches, the peak of aperiodic mode is located at lower wave-numbers. 
Moreover, the effect of suprathermal electrons, lowering the unstable wave-numbers is again
evident. 

Of these two branches of EFH instabilities, the aperiodic branch (A-EFH) is by far the fastest 
growing and may therefore be the most effective in the relaxation of temperature anisotropy. 
Figure \ref{f5} provides a better visualization of its properties by mapping with contour 
plots the maximum growth rate as a function of the wave number $kc/\omega_{p,e}$ and the angle 
of propagation $\theta$. Panel (a) displays the solutions in the presence of suprathermal electrons 
($\kappa=4$) with maximum growth rates (color bar) peaking at $\gamma_m/ |\Omega_e|\approx0.095$ 
(for $\theta_m = 72^\circ$), markedly higher than the peak of Maxwellian solutions at $\gamma_m/ 
|\Omega_e|\approx0.065$ (for $\theta_{m}=65^\circ$) in panel (b). For both cases peaks are 
obtained at $k_m\approx0.55~\omega_{p,e}/c$. These results show clearly that A-EFH instability 
is stimulated in the presence of suprathermal electrons, 
and the fastest growing mode develops at a higher angle of propagation.

\subsection{The influence of suprathermal electrons on A-EFH thresholds}

\begin{figure}
\includegraphics[scale=0.59]{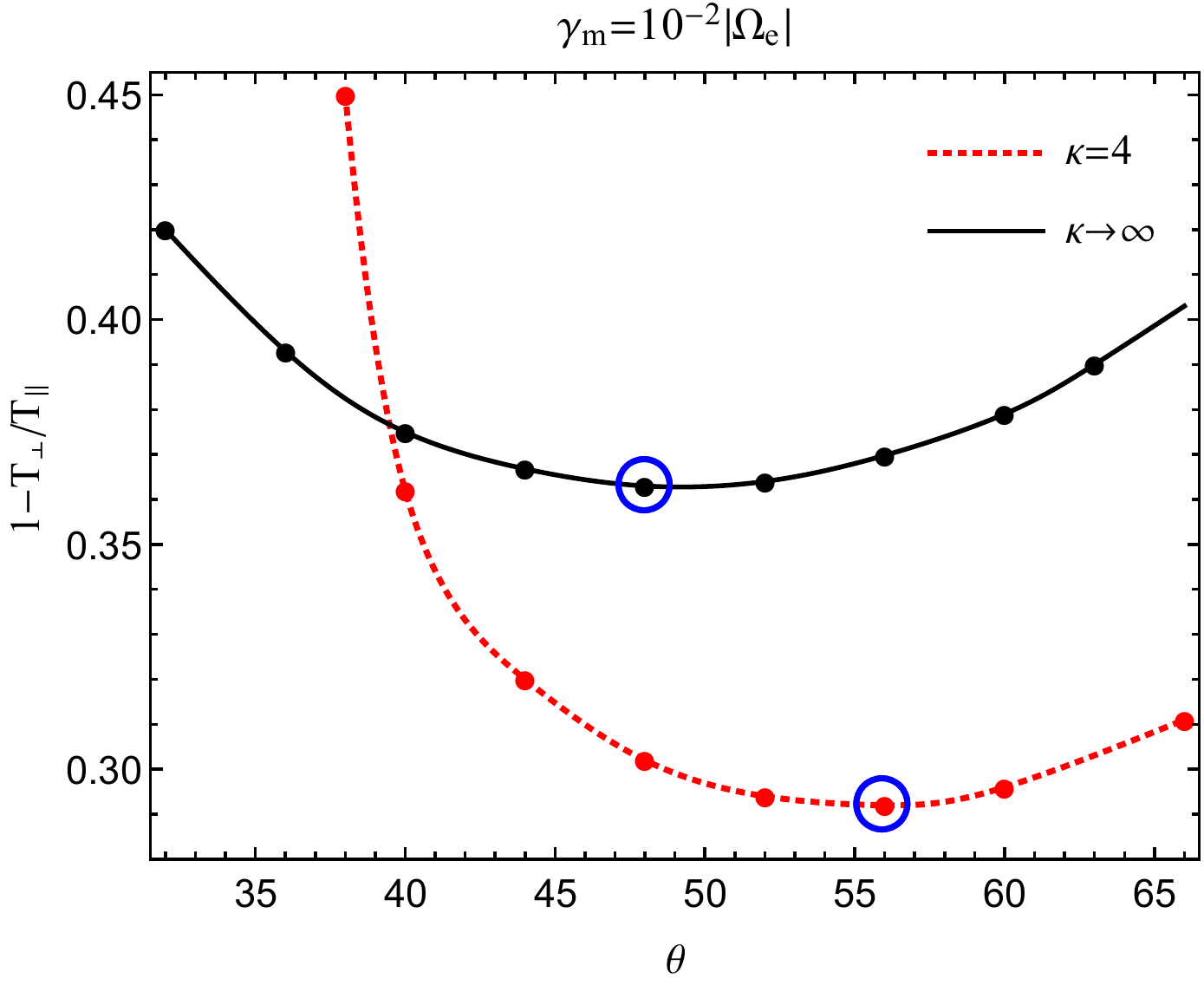}
\caption{Complementary anisotropy ($1-T_\perp/T_\parallel$) thresholds
for maximum growth rates $\gamma_m=10^{-2}|\Omega_e|$ of the A-EFH instability, as a function of 
$\theta$, for $\kappa=4$ (red dotted line) and $\kappa\rightarrow \infty$ (black solid line). Blue 
open circles indicate the angles $\theta_m$ corresponding to the lowest anisotropy. }\label{f6}
\end{figure}

\begin{figure}
\includegraphics[scale=0.58]{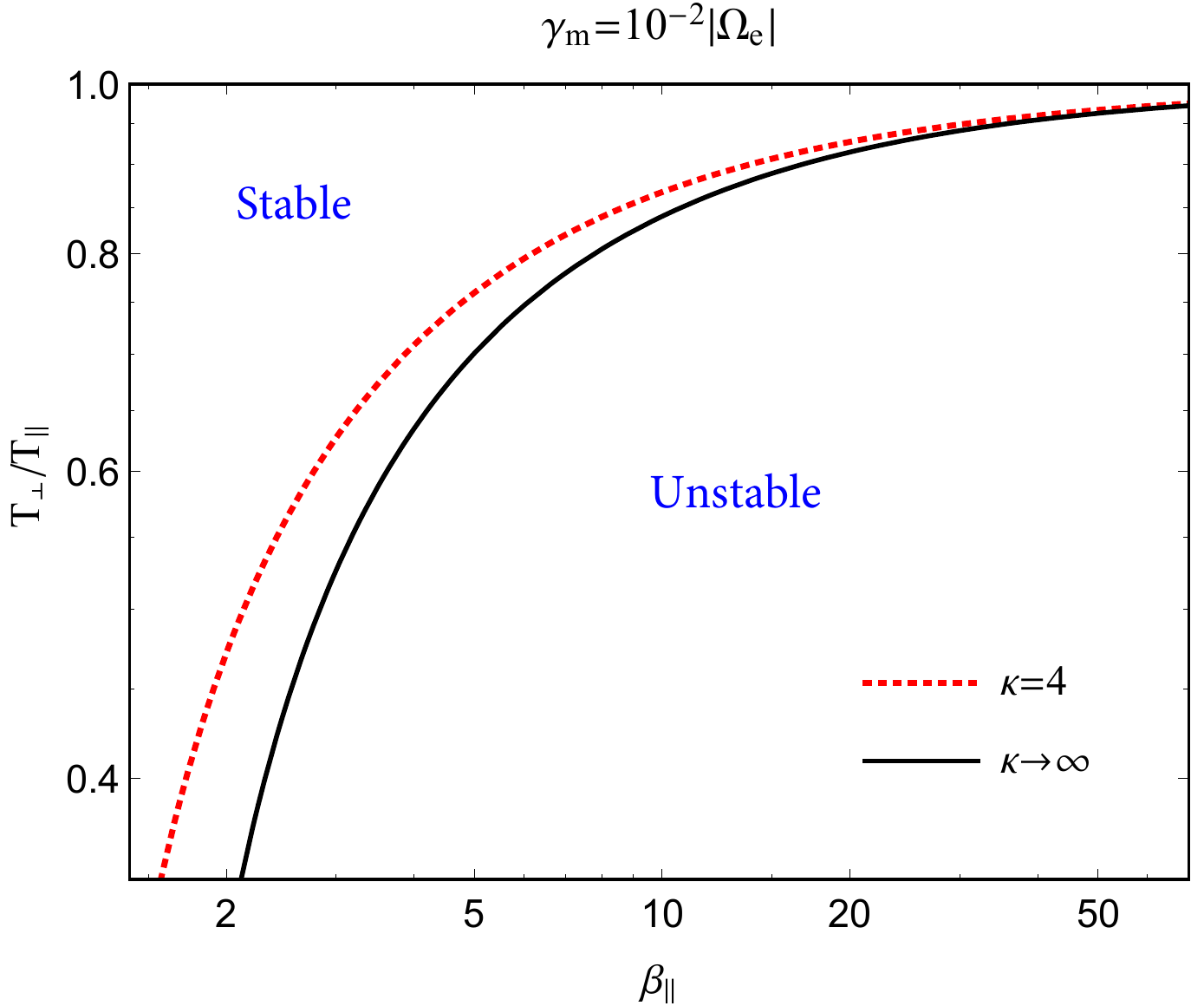}
\caption{Electron anisotropy threshold (maximum growth rates $\gamma_m=10^{-2}|\Omega_e|$)
for A-EFH instability as a function of parallel plasma beta $\beta_\parallel$ for 
$\kappa=4$ (red dotted line) and $\kappa\rightarrow \infty$ (black solid line).}\label{f7}
\end{figure}

In this section we describe the anisotropy thresholds of the aperiodic branch identified as the 
fastest growing mode. These thresholds characterize the unstable plasma conditions for the 
lower levels of maximum growth rates, approaching marginal condition of stability ($\gamma_m/
|\Omega_e|\rightarrow 0$). Here we compare the anisotropy thresholds derived for a maximum
growth rate $\gamma_m=10^{-2}|\Omega_e|$. Figure \ref{f6} displays these isocontours as 
a function of $\theta$ and complementary anisotropy $1-T_\perp/T_\parallel$ for $\kappa=4$ 
(red) and Maxwellian limit ($\kappa \to \infty$, black). These thresholds show a non-monotonous 
variation, decreasing and then increasing with increasing $\theta$. In the presence of 
suprathermals ($\kappa=4$) the lowest (minimum) anisotropy $(1-T_\perp/T_\parallel)_m = 0.292$ 
is found at $\theta_m \simeq 56^\circ$ (opened circle on red line), and is much lower than that 
obtained for Maxwellian limit, i.e., $(1- T_\perp/T_\parallel)_m = 0.36 $ at $\theta_m \simeq 
48.5^\circ$ (opened circle on black line). 
Evaluating the lowest (threshold) anisotropies for different values of electron (parallel) plasma 
beta $\beta_\parallel = 2,3, 4, 5, 6, 8, 20, 50$, enabled us to determine the general instability 
thresholds, as a function of plasma parameters, e.g., $\beta_\parallel$. 
These thresholds are displayed in Figure \ref{f7} using temperature anisotropy introduced in 
Table~1, i.e., $T_\perp /T_\parallel$ (as also used in recent studies to facilitate comparison 
with the observations). As a function of $\beta_\parallel$, this threshold is fitted to an inverse 
correlation law  \citep{Gary2003}
\begin{align}
\frac{T_\perp}{T_\parallel}=1-\frac{a}{\beta_\parallel^b}\,,
\end{align}
where for the fitting parameters we find $(a, b)=(0.96, 0.86)$ in the presence of suprathermal 
electrons ($\kappa=4$, dotted-red), and $(a, b) = (1.27, 0.90)$ for Maxwellian limit ($\kappa 
\rightarrow \infty$, solid-black). The instability thresholds are determined for an extended 
range of electron plasma beta $1<\beta_\parallel<70$, to include conditions specific to various
plasma conditions in heliophere, e.g., solar flares, solar wind or planetary magnetospheres
\citep{Stverak2008}. 

Suprathermal electrons have a stimulating effect on the instability, diminishing 
the threshold to lower values of $\beta_\parallel$ and closer to isotropy ($A \to 1$), 
as shown in Figure~\ref{f7}. The unstable regime expands confirming results in 
Figures \ref{f5} and \ref{f6}. This stimulation is indeed more pronounced at low 
values of $\beta_\parallel$, but decreases by increasing this parameter.

%

\section{Conclusions}

In-situ measurements of the solar wind electrons suggest that their velocity distributions 
exhibit power-law tails well described by the bi-Kappa functions \citep{Vasyliunas1968, 
Pierrard2010, Stverak2008}. Recent studies show that suprathermal electrons, if correctly 
interpreted, can significantly change the (linear) dispersive characteristics of kinetic 
instabilities \citep{Vinas2015, Lazar2015, Shaaban2016, Lazar2017b, Shaaban2017, 
Shaaban2018b}. However, these studies are in general restricted to modes propagating parallel 
to the background magnetic field, while the obliquely propagating unstable modes are rarely 
investigated and only for idealized (bi-)Maxwellian plasmas. In the present work we have 
analyzed the full spectrum of firehose instabilities triggered by the anisotropic electrons 
in the presence of their suprathermal populations. The general linear dispersion relations 
have been solved for arbitrary angles of propagation using an advanced numerical solver 
dedicated to plasmas with bi-Kappa populations, named DSHARK \citep{Astfalk2015}. 

In Section \ref{sec:2} we have described both the periodic and aperiodic branches 
of the electron firehose unstable solutions for a set of plasma parameters typically 
experienced in the solar wind. In the oblique directions the periodic branch extends 
only to low angles, while the aperiodic firehose instability develops with much 
higher growth rates at highly oblique directions. After a detailed description 
in Figures \ref{f1}--\ref{f3}, which teaches us how to differentiate between these
two branches, in Figure \ref{f4} we have shown that suprathermal electrons stimulate
both branches, increasing the growth rates of the fastest growing modes as indicated 
by the peaking (maximum) growth rates. The aperiodic firehose remains the fastest 
growing and Figure \ref{f5} provides a complete picture of the influence of 
suprathermal electrons on its peaking growth rates, as this depends on the wave-number 
and the propagation angle.
The anisotropy thresholds in Figures \ref{f6} and \ref{f7} provide an
important indication about conditions limiting the instability of aperiodic firehose
mode. Particularly important are the general thresholds provided in Figure 
\ref{f7}, which are derived exclusively in terms of plasma parameters, and 
show that conditions favorable to this instability may significantly extend
in the presence of suprathermal electrons. 

To conclude, we have shown that suprathermal electrons present in space plasmas,
and in particular in the solar wind and planetary environments, have an important 
impact stimulating both branches of the firehose instability, of which the aperiodic 
firehose remains dominant, and may induce new regimes of instability at lower
wavenumbers. Comparing to standard results for bi-Maxwellian plasmas, which 
ignore the effects of suprathermal electrons, our results predict an enhanced 
role of firehose instabilities, especially in collision-poor plasmas from 
space where these instabilities can prevent the large deviations from isotropy
and, thus, explain the observations. These new regimes of firehose 
instability may also be relevant for many other scenarios in astrophysics which assume 
an implication of this instability, e.g., solar flares, sites of magnetic field 
reconnection, accretion flows or plasma jets leading to shocks and co-rotating 
interactions in heliosphere, interstellar medium and galaxy clusters.

%
\begin{table}
	\centering
	\caption{Plasma parameters set used in the present study}
   \label{t1}
	\begin{tabular}{lccc} 
		\hline
		 & Electrons ($i=e$) &  Protons ($i=p$)\\
		\hline
		$\beta_{i,\parallel}$ & 4 & 4\\
		$\beta_{i,\perp}$     & 2 & 4\\
		$A_{i}=T_{i,\perp}/T_{i,\parallel}$ & 0.5 & 1\\
		$m_p/m_i$        &   1836 & 1.0 \\
		$\kappa$    & 4, $\infty$ & $\infty$\\
		\hline
	\end{tabular}
\end{table}

\section*{Acknowledgements}

These results were obtained in the framework of the projects SCHL 201/35-1 (DFG-German 
Research Foundation), GOA/2015-014 (KU Leuven), G0A2316N (FWO-Vlaanderen), and C 90347 
(ESA Prodex 9). S.M. Shaaban acknowledges support by a FWO Postdoctoral Fellowship (Grant 
No. 12Z6218N), and a FWO Travel grant for long stay abroad (Grant No. V419818N). 
Thanks are due to Patrick Astfalk for valuable discussions and for providing the 
DSHARK code. We also appreciate the support from the International Space Science Institute 
(ISSI) for hosting the international ISSI team on Kappa Distributions, which triggered 
fruitful discussions that were beneficial for the work presented here.



\bibliographystyle{mnras}
\bibliography{papers} 


\appendix 
\section{Elements of the dielectric tensor}
\label{sec:appendix}
The general linear dispersion relations for the plasma electromagnetic modes propagating at an arbitrary angle with respect to the uniform background magnetic field, $\bm{B}_0=B_0 \hat{e}_z$, in a bi-kappa distributed plasma is given by \citep{Summers1994, Shaaban2018St}
\begin{align}
0 = \det D(\omega, k, \theta)\,,
\end{align}
with
\begin{align}
D(\omega, k&, \theta) =\nonumber\\
&\left|\begin{array} {ccc} 1-\frac{c^2 k^2_\parallel}{w^2}+ \epsilon_{xx}& \epsilon_{xy} & 
\frac{c^2 k_\parallel k_\perp}{w^2} +  \epsilon_{xz} \\ \\
\epsilon_{yx} & 1-\frac{c^2 k^2}{w^2} +  \epsilon_{yy} & \epsilon_{yz} \\ \\
 \frac{c^2 k_\parallel k_\perp}{w^2} +  \epsilon_{zx} &  \epsilon_{zy} &
1-\frac{c^2 k^2_\perp}{w^2} +  \epsilon_{zz} \end{array} \right|,
\label{eq:A1}
\end{align}

The dielectric elements in Eq.(\ref{eq:A1}) can be expressed in terms of the modified plasma dispersion function $Z_\kappa (\xi_j)$ and the Bessel functions of the first kind  $J_n(\mu)$ as follows 
\begin{align}
\epsilon_{xx} =&\sum_j  \sum_{n=-\infty}^{n=+\infty}
{\omega_{p, j}^2 \over \omega^2} \frac{n^2}{\lambda^2_j}\left[ \Lambda_j K_1 \int_{0}^{\infty} 
\frac{\mu J_n^{2}(\mu) d\mu}{(1+\mu^2/2\lambda~\kappa)^{\kappa+3/2}}\right.\nonumber \\ 
&\left.+\left(\Lambda_j~\xi_j+\dfrac{\omega}{k_\parallel \Theta_{\parallel,j}}\right) K_2 \int_{0}^{\infty} \frac{\mu J_n^{2}(\mu) d\mu}{(1+\mu^2/2\lambda~\kappa)^{\kappa+2}}\right.\nonumber \\
&\left. \times Z_{\kappa+1}\left(\xi_j\sqrt{\frac{\kappa+1}{1+\mu^2/2\lambda}}\; \right)d\mu\right]
\end{align}
\begin{align}
\epsilon_{yy} =&\sum_j  \sum_{n=-\infty}^{n=+\infty}
{\omega_{p, j}^2 \over \omega^2\lambda^2_j}\left[ \Lambda_j ~K_1 \int_{0}^{\infty} 
\frac{\mu^3 J'^2_n(\mu) d\mu}{(1+\mu^2/2\lambda~\kappa)^{\kappa+3/2}}
\right.\nonumber \\
&\left. +\left(\Lambda_j~\xi_j+\dfrac{\omega}{k_\parallel \Theta_{\parallel,j}}\right)  K_2 \int_{0}^{\infty} \frac{\mu^3 J'^2_n(\mu) d\mu}{(1+\mu^2/2\lambda~\kappa)^{\kappa+2}}\right.\nonumber \\
&\left. \times Z_{\kappa+1}\left(\xi_j\sqrt{\frac{\kappa+1}{1+\mu^2/2\lambda}}\; \right)d\mu\right]
\end{align}
\begin{align}
\epsilon_{zz} =&\sum_j  \sum_{n=-\infty}^{n=+\infty}
{\omega_{p, j}^2 \over \omega^2}{\Theta_{\parallel, j}^2 \over \Theta_{\perp, j}^2}\frac{2~\xi_j}{\lambda^{3/2}}\left(\Lambda_j~\xi_j+\dfrac{\omega}{k_\parallel \Theta_{\parallel,j}}\right) \nonumber\\
&\times\left[ K_1 \int_{0}^{\infty} \frac{\mu J^2_n(\mu) d\mu}{(1+\mu^2/2\lambda~\kappa)^{\kappa+3/2}}+\xi_j K_2 \int_{0}^{\infty} \frac{\mu J^2_n(\mu) d\mu}{(1+\mu^2/2\lambda~\kappa)^{\kappa+2}} \nonumber\right. \\
& \left. \times Z_{\kappa+1}\left(\xi_j \sqrt{\frac{\kappa+1}{1+\mu^2/2\lambda}}\; \right)d\mu\right]
\end{align}
\begin{align}
\epsilon_{xy} =-\epsilon_{yx}=&\sum_j  \sum_{n=-\infty}^{n=+\infty}
{\omega_{p, j}^2 \over \omega^2} \frac{i n}{\lambda^2_j}\left[ \Lambda_j K_1 \int_{0}^{\infty} 
\frac{\mu^2 J_n(\mu)J'_n(\mu) d\mu}{(1+\mu^2/2\lambda~\kappa)^{\kappa+3/2}}\right.\nonumber \\
&\left. +\left(\Lambda_j~\xi_j+\dfrac{\omega}{k_\parallel \Theta_{\parallel,j}}\right) K_2 \int_{0}^{\infty} \frac{\mu^2 J_n(\mu)J'_n(\mu) d\mu}{(1+\mu^2/2\lambda~\kappa)^{\kappa+2}}\right.\nonumber \\
&\left. \times Z_{\kappa+1}\left(\xi_j\sqrt{\frac{\kappa+1}{1+\mu^2/2\lambda}}\;  \right)d\mu\right]
\end{align} 
\begin{align}
\epsilon_{xz}& =\epsilon_{zx}\nonumber=\sum_j  \sum_{n=-\infty}^{n=+\infty}
{\omega_{p, j}^2 \over \omega^2}{\Theta_{\parallel, j} \over \Theta_{\perp, j}}
\frac{\sqrt{2}n}{\lambda^{3/2}}\left(\Lambda_j~\xi_j+\dfrac{\omega}{k_\parallel \Theta_{\parallel,j}}\right)\\
&\times\left[ K_1 \int_{0}^{\infty} \frac{\mu J^2_n(\mu) d\mu}{(1+\mu^2/2\lambda~\kappa)^{\kappa+3/2}}+ \xi_j K_2 \int_{0}^{\infty} \frac{\mu J^2_n(\mu) d\mu}{(1+\mu^2/2\lambda~\kappa)^{\kappa+2}} \right. \nonumber\\
& \left. \times Z_{\kappa+1}\left(\xi_j\sqrt{\frac{\kappa+1}{1+\mu^2/2\lambda}}\; \right)d\mu\right]
\end{align}
\begin{align}
\epsilon_{yz} &=-\epsilon_{zy}=\sum_j  \sum_{n=-\infty}^{n=+\infty}
{\omega_{p, j}^2 \over \omega^2}{\Theta_{\parallel, j} \over \Theta_{\perp, j}}
\frac{-\sqrt{2}i}{\lambda^{3/2}}\left(\Lambda_j~\xi_j+\dfrac{\omega}{k_\parallel \Theta_{\parallel,j}}\right)\nonumber \\
& \times \left[ K_1 \int_{0}^{\infty} \frac{\mu^2 J_n(\mu)J'_n(\mu) d\mu}{(1+\mu^2/2\lambda~\kappa)^{\kappa+3/2}}+ \xi_j K_2 \int_{0}^{\infty} \frac{\mu^2 J_n(\mu)J'_n(\mu) d\mu}{(1+\mu^2/2\lambda~\kappa)^{\kappa+2}}\nonumber \right. \\
& \left. \times Z_{\kappa+1}\left(\xi_j\sqrt{\frac{\kappa+1}{1+\mu^2/2\lambda}}\; \right)d\mu\right]
\end{align}
where $j$ denotes different plasma species, e.g., electrons (subscript $j=e$), protons ($j=p$), $\omega_{p, j}=\sqrt{4\pi n_j e^2/m_a}$ and $\Omega_j=e B_0/m_j c$ are the
non-relativistic plasma frequency and the gyro-frequency of species $j$, respectively, 
\begin{align*}
K_1=\dfrac{\kappa^2-1/4}{\kappa^2}, \; \; \; K_2=\frac{2\kappa-1}{2\kappa} \left(\frac{\kappa+1}{\kappa}\right)^{3/2},
\end{align*}
\begin{align*}
\Lambda_j=A_j-1, \;\; \lambda_j = {k_\perp^2 \Theta_{\perp, j}^2 \over 2~ \Omega_j^2}, \;\;  \xi_j = {\omega - n \Omega_j \over k_\parallel \Theta_{\parallel, j}}
\end{align*}
and $Z_\kappa(\xi_j)$ is the plasma modified dispersion function given by \citep{Summers1994}
\begin{align*}
Z_\kappa\left( \xi_{j}\right) =&\frac{\pi ^{-1/2}}{\kappa_{j}^{3/2}}\frac{\Gamma \left( \kappa_{j} \right) }{\Gamma \left(\kappa_{j} -1/2\right) }\int_{-\infty }^{\infty }\frac{\left(1+x^{2}/\kappa_{j} \right) ^{-\kappa_{j}-1}}{x-\xi_{j}}dx,\ \  \Im \left(\xi_{j}\right) >0.
\end{align*}
%

\bsp	
\label{lastpage}
\end{document}